%% file: long.tex
\documentstyle[psfig,12pt]{article}
\input lmacros

\def\comment#1{  \begin{raggedright}{\tt [#1]}\end{raggedright}}
\begin{document}
\baselineskip=14.5pt \pagestyle{plain} \setcounter{page}{1}
\renewcommand{\thefootnote}{\fnsymbol{footnote}}
\newcommand{\da}{\dot{a}}
\newcommand{\db}{\dot{b}}
\newcommand{\dn}{\dot{n}}
\newcommand{\dda}{\ddot{a}}
\newcommand{\ddb}{\ddot{b}}
\newcommand{\ddn}{\ddot{n}}
\newcommand{\pa}{a^{\prime}}
\newcommand{\pb}{b^{\prime}}
\newcommand{\pn}{n^{\prime}}
\newcommand{\ppa}{a^{\prime \prime}}
\newcommand{\ppb}{b^{\prime \prime}}
\newcommand{\ppn}{n^{\prime \prime}}
\newcommand{\fda}{\frac{\da}{a}}
\newcommand{\fdb}{\frac{\db}{b}}
\newcommand{\fdn}{\frac{\dn}{n}}
\newcommand{\fdda}{\frac{\dda}{a}}
\newcommand{\fddb}{\frac{\ddb}{b}}
\newcommand{\fddn}{\frac{\ddn}{n}}
\newcommand{\fpa}{\frac{\pa}{a}}
\newcommand{\fpb}{\frac{\pb}{b}}
\newcommand{\fpn}{\frac{\pn}{n}}
\newcommand{\fppa}{\frac{\ppa}{a}}
\newcommand{\fppb}{\frac{\ppb}{b}}
\newcommand{\fppn}{\frac{\ppn}{n}}
\newcommand{\A}{A}
\newcommand{\B}{B}
\newcommand{\mmu}{\mu}
\newcommand{\mnu}{\nu}
\newcommand{\ii}{i}
\newcommand{\jj}{j}
\newcommand{\jl}{[}
\newcommand{\jr}{]}
\newcommand{\ml}{\sharp}
\newcommand{\mr}{\sharp}



\begin{titlepage}

\begin{flushright}
MIT-CTP-3146 \\
hep-th/0105132
\end{flushright}
\vfil

\begin{center}
{\huge Open and Closed String Interpretation of SUSY CFT's on
Branes with Boundaries
}\\[8pt]
\end{center}

\vfil
\begin{center}
{\large Andreas Karch and Lisa Randall}
\end{center}

$$\seqalign{\span\TL & \span\TT}{
 & Center for Theoretical Physics,
  \cr\noalign{\vskip-1.5\jot}
   & Massachusetts Institute of Technology, Cambridge, MA  02139-4307, USA
  \cr\noalign{\vskip0.5\jot}
}$$ \vfil

\begin{center}
{\large Abstract}
\end{center}

\noindent

We consider certain supersymmetric configurations of intersecting
branes and branes ending on branes and analyze the duality between
their open and closed string interpretation. The examples we study
are chosen such that we have the lower dimensional brane realizing
an $n+1$ dimensional conformal field theory on its worldvolume and
the higher dimensional one introducing a conformal boundary.
We also consider two CFTs, possibly with different central
charges, interacting along a common conformal boundary. 
We show with a probe
calculation that the  dual closed string description is in terms
of gravity in an AdS$_{n+2}$ bulk with an AdS$_{n+1}$ defect
or two different AdS$_{n+2}$ spaces joined along a defect.
We also comment briefly on the expected
back-reaction.

\vfil
\begin{flushleft}
July 2000
\end{flushleft}

{\vskip 5pt \footnoterule\noindent {\footnotesize $\,^*$\ {\tt
karch@mit.edu}, {\tt randall@mit.edu}
}}
\end{titlepage}
\newpage
\baselineskip=15.5pt

\section{Introduction}

AdS/CFT duality can be understood as the statement that a given
brane configuration in string theory has two equivalent
descriptions, one in terms of open and one in terms of closed
strings. The examples in the original work of \cite{maldacena} are
branes that in a certain decoupling limit realize a supersymmetric
$n+1$ dimensional conformal field theory on their worldvolume, the
open string description. In the dual language, the theory is
described in terms of closed strings propagating in the
near-horizon geometry of the corresponding black brane solution.
The conformal invariance of the field theory is reflected in the
fact that the near horizon geometry is AdS$_{n+2}$.

In \cite{kr} it was argued that the dual description of a brane
with an AdS$_{n+1}$ inside that AdS$_{n+2}$, is a conformal field
theory with boundary or, more generally, with a codimension one
defect. In order to verify this claim we study supersymmetric
configurations in string theory, with branes intersecting in such
a way that the lower dimensional one is one of the basic CFT
examples of \cite{maldacena}, while the higher dimensional one
appears as a supersymmetric codimension one defect in this CFT,
such that the lower dimensional brane can end. In particular we
discuss the D3 brane ending on 5-branes in IIB string theory, D1
D5 black strings ending on D3 branes and M2 branes ending on M5
branes.

The choice of examples is motivated by the requirement that in the
decoupling limit the open strings describe a supersymmetric CFT on
a manifold with boundary, that is one half of our duality
conjecture. In order to establish that the dual closed string
description involves AdS$_{n+1}$ branes in AdS$_{n+2}$ we show
that in the probe approximation the higher dimensional brane
indeed spans an AdS worldvolume in the near horizon geometry of
the lower-dimensional brane.

This scenario should be  contrasted with the case of a localizing
Minkowski brane. It has been believed for a while that the
standard Minkowski$_{d+1}$ brane is dual to introducing a UV
cutoff in the field theory, breaking the conformal group from
$SO(d+1,2)$ to the $d+1$ dimensional Lorentz group preserved by
the brane. This cutoff does not correspond to any known field
theory cutoff, rather one should view the supergravity description
as the definition of this new cutoff with somewhat unusual
properties. Therefore quantitative comparison is difficult;
however many qualitative features can be explained nicely, see
e.g. \cite{lisa,rattazzi}. In the AdS case we impose a spatial
instead of a momentum cutoff, yielding a dual theory with a well-defined
lagrangian description.

In the next section, we  review the set-up and  duality of
\cite{kr}. In Section 3, we give some general arguments about how
a probe 5-brane is embedded in the near horizon geometry created
by the D3 branes. In Section 4, we generalize this result to M2
branes ending on M5 branes and black strings ending on D3s. For the
latter one we compare with the recent probe calculation of Bachas
and Petropoulos \cite{bp}. In Section 5, we perform a Born-Infeld
probe calculation for the 4d case, verifying the results obtained
in Section 3. In Section 6,  we consider the theory beyond the
probe approximation and study how the interplay of bulk versus
boundary modes of the CFT is reflected in the gravity
fluctuations. We conclude in the final section.

\section{The Supergravity Dual for a CFT on the Half-Plane}

\begin{figure}
   \centerline{\psfig{figure=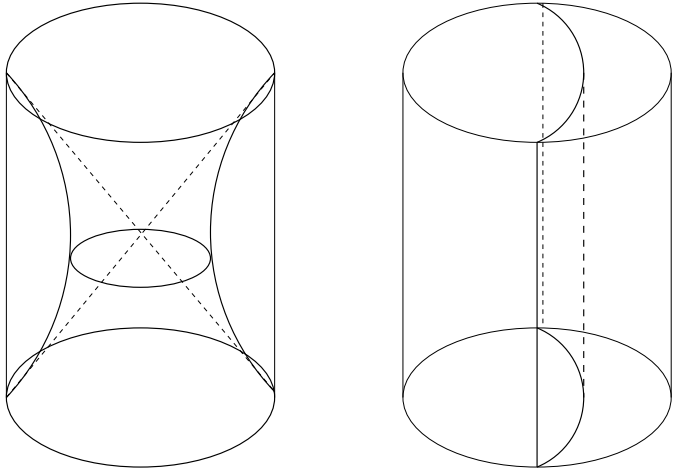,width=3.5in}}
    \caption{Global embedding of the dS$_2$ and AdS$_2$
branes in AdS$_3$ respectively. The dS case is only
included for comparison.}
\label{emb}
 \end{figure}

The basic observation giving rise to the duality of \cite{kr} is
the embedding of the AdS$_{d+1}$ brane in AdS$_{d+2}$, as it is
depicted in Fig.\ref{emb}. The big cylinder represents the
AdS$_{d+2}$. The AdS$_{d+1}$ brane has the same global time as the
big AdS, that is: it does not move. It ends at the equator of the
S$^d$ that is the boundary of the AdS$_{d+2}$, effectively cutting
the boundary in half. As we increase the tension the brane curves
more and more into the bulk, approaching the true boundary. It
nevertheless always ends on the equator of the sphere. For a
positive tension brane the part of the cylinder to the right of
the brane in Fig.\ref{emb} is removed, including half of
the ``true'' boundary. Gravity in the remaining bulk space-time
should have a holographic description in terms of the CFT on the
disk plus a theory living on the brane. Appealing to AdS$_{d+1}$
holography one can reduce the description further to a CFT living
on the disk with a boundary action living on the common boundary
of the disk that remained of the ``true'' boundary and the brane,
basically specifying the boundary conditions for the CFT.
Observing that the $SO(d,2)$ symmetry group preserved by the
AdS$_{d+1}$ brane is precisely the subgroup of the conformal group
leaving a boundary invariant \cite{cardy}, one is led to believe
that the holographic dual of the gravity set-up we described is
indeed the original CFT together with a set of conformal boundary
conditions specified by the properties of the brane.

If one is looking at the Poincare patch of the above solution
only, the original dual CFT lives on Minkowski space
instead of the sphere, and after introducing the AdS brane
in the bulk, similarly the boundary CFT will now live on
half of Minkowski space, the ``upper half plane''. Again
the boundary conditions reflect the properties of the
brane. If one does not impose an orbifold condition
relating the independent fluctuations in the
two halves of space-time, one ends up with 2 CFTs, each
defined on half of Minkowski space, interacting along their common
boundary. Or in other words, a CFT living on all of space with a
codimension one conformal defect.
In general one can couple two different CFTs even with different
central charge along a boundary in this fashion by
having the background curvature of the AdS$_{d+2}$
jump across the brane. The string theory
setup we are describing for non-zero $q$ gives a supersymmetric
realization of such a system.

\section{The Probe Picture}

\subsection{The Embedding from AdS/CFT}

In the introduction we  argued that intersecting D3
D5 branes or similar set-ups that are stringy realizations of
a conformal field theory on a manifold with boundary should
have a near horizon geometry that looks like a warped
product of AdS$_4$ with a gravity localizing warp factor
times a compact internal space.
In this section we will show that when neglecting the backreaction
of the D5 branes on the geometry, one indeed finds that they
wrap an equatorial $S^2$ inside the $S^5$ and live on an AdS$_4$
inside the AdS$_5$. The AdS$_4$ can be made arbitrarily flat
by having some of the D3 branes end on the D5 brane.
Let us start with the case of intersecting branes. That is we have
branes along
\begin{center}
\vspace{.2cm}
\begin{tabular}{|c||c|c|c|c|c|c|c|c|c|c|}
\hline
&$x^0$&$x^1$&$x^2$&$x^3$&$x^4$&$x^5$&$x^6$&$x^7$&$x^8$&$x^9$\\
\hline
D 5&x&x&x&x&x&x&o&o&o&o\\
\hline
D 3&x&x&x&o&o&o&x&o&o&o\\
\hline
\end{tabular}
\vspace{.2cm}
\end{center}
The near horizon geometry of the $N$ D3 branes is AdS$_5$ $\times$ $S^5$.
When the metric is written as
\eqn{adsmetric}{
ds^2 =L^2 \left ( \; - du^2/u^2 +
\frac{u^2}{k_N^2} (dt^2 - dx^2 - dy^2 - dz^2) + ds^2_{S5}
\; \right )
}
where we define $k_N$ as
\eqn{kn} {k_N = \sqrt{ 4 \pi g_s N}}
and the curvature length is given by
\eqn{5dlength}{L^2 = l_s^2 \sqrt{ 4 \pi g_s N} = k_N l_s^2,  }
one can make the following identifications \cite{maldacena}
between the coordinates of the near horizon region and
the coordinates describing the flat embedding coordinates of the branes:
\begin{center}
\vspace{.2cm}
\begin{tabular}{r c l}
 $u^2$    &  $\rightarrow$ & $\frac{1}{l_s^4} (x_3^2 + x_4^2 + x_5^2 + x_7^2
+
x_8^2 + x_9^2)$  \\
 $x$  &  $\rightarrow$ & $x_6$ \\
 $t$, $y$, $z$    &  $\rightarrow$ & $x_0$, $x_1$, $x_2$ \\
\end{tabular}
\vspace{.2cm}
\end{center}
and the $S^5$ is the sphere of radius $L$ surrounding the D3 brane,
\eqn{thesphere}{S^5 \; \; : \; \; x_3^2 + x_4^2 + x_5^2 + x_7^2 + x_8^2 +
x_9^2 = L^2.}
Of course one can always change coordinates to $u = k_N v$,
such that the metric takes the standard AdS form
\eqn{adsmetric2}{
ds^2 =L^2 \left ( \; - dv^2/v^2 +
v^2 (dt^2 - dx^2 - dy^2 - dz^2) +  ds^2_{S5}
\; \right ) .
}

Now if we add a SUSY probe D5 brane, we know it will span the
wordvolume given by \eqn{thefivebrane}{x_6=x_7=x_8=x_9=0.} That
is, it is wrapping the radius $L$ $S^2$ inside the $S^5$ given by
\eqn{twosphere}{x_3^2+x_4^2+x_5^2 =L^2} and is stretching along
the surface $x=0$ inside the AdS$_5$.
The apparent tachyonic instability in fact is above
the Breitenlohner-Freedman bound \cite{BF}, as we will
show later. Note that the induced metric
on any surface of the form \eqn{smallads}{ x= \frac{C}{v} } is an
AdS$_4$ written in the same coordinate system with curvature
length \eqn{smalll}{l^2 = L^2 (1 + C^2) .} When transformed into
coordinates covering global AdS, one again sees  this hypersurface
is the one described in Fig.\ref{emb}, where $C$ is the minimal
distance between the AdS brane and the center of the cylinder, see
\cite{kr,bp}. For D3 branes intersecting the D5, the D5 brane
worldvolume is the straight (that is, in Fig.\ref{emb} it just
cuts straight through the middle of AdS) but maximally curved
AdS$_4$ with curvature radius $l=L$.

Of course this is just the probe approximation. Ultimately, one has
to account for the back reaction. Before we do this, it is also of
interest to study the case where a large number $q$ of the $N$ D3
branes end on the D5 brane. It will turn out this generates a
large separation between $l$ and $L$, corresonding to nonzero $C$.

So now we consider the situation where
 $q$ of the $N$ D3 branes
end on the D5 brane. Obviously this requires $q\leq N$. The branes
now bend as in \cite{witten4d}. The bending is the solution to a
minimal area Laplace problem \cite{witten4d}. As a reaction to
being pulled on by the D3 branes, the D5 brane position along
$x_6$ becomes a function of $r^2= x_3^2 + x_4^2 + x_5^2$. Since
this is codimension 3, the bending is of the form
\eqn{bending}{x_6 = l_s^2 \; \frac{\tilde{C}}{r}.} Using that
along $x_7=x_8=x_9=0$ one just gets $r=l_s^2 u$, \bending\
translates into \eqn{into}{x= \frac{\tilde{C}}{u}
 = \frac{\tilde{C}}{k_N v} = \frac{C}{v}}
and the bent
D5 brane now once more lives along an AdS$_4$ subspace of AdS$_5$,
this time given by nonzero $C$. In order to determine $C$,
note that $\tilde{C}=C k_N$ is proportional to the ratio of the
tension of the brane that is
pulling to the tension of the brane that is being pulled. For $N$ D3
branes pulling on one side and $N-q$ D3 brane pulling on the
other side of $M$ D5 branes we therefore get
\eqn{c}{ C = \alpha \;  \;
 \frac{q T_{D3}}{M T_{D5} } \; \frac{1}{l_s^2 k_N} }
where $\alpha$ contains the factors of 2 and $\pi$
and will be determined later. So we see that for large $q/k_N$,
we get a brane with large curvature radius.

\section{Other Examples of the Geometry of Intersecting Branes}

Although we focus on the D3-D5 brane system, which can
result in localized four-dimensional gravity, a similar
analysis can be applied to other systems of branes. We describe
a few below.  One example has actually already
been studied \cite{bp}
and is consistent with our treatment. In other cases,
we learn about the expected supergravity solution for
other string constructions.

\subsection{6d Black String with RR charges}

We first consider a 6d black string ending on a D3 brane. The
black string can either be made out of D1 D5 strings or F1 NS5
strings, where the $Q_5$ 5-branes are always wrapped on the
internal 4 manifold $M_4$ that one used to compactify to 6
dimensions. The role of the D5 brane from above is now played by a
D3 brane. Note that it is only the $Q_1$ strings that can actually
end on the D3 brane; the 5-brane part of the black string is not
allowed to do so since the 5-brane wraps the internal manifold,
while the D3 branes are a point on the internal manifold. So this
time we have $Q_1$ strings and $Q_5$ 5-branes on one side of $M$
D3 branes and $Q_1 -q$ and $Q_5$ on the other side. In the D1 D5
system the 3d radius of curvature $L^2$ goes like \eqn{d1d5}{ L^2
= g_{s,6d} \sqrt{Q_1 Q_5} l_s^2} so it will jump by $q$ units as
above. A string ending on a 3-brane is also codimension 3, so the
analysis of the previous section goes through unchanged, we end up
with an AdS$_2$ $\times$ $S^2$ inside the AdS$_3$ $\times$ $S^3$
set up by the black string with $q$ units of worldvolume gauge
field flux and the AdS$_2$ has a radius of curvature given by
$l^2= L^2 (1+C^2)$ with $C$ as in \c\ with D5 and D3 branes
replaced by D3 and D1 branes respectively and \eqn{kfordone}{k_N =
g_{s,6d} \sqrt{Q_1 Q_5}.}

\subsection{6d Black String with NSNS charges}

For the system with F1 NS5 branes, the 3d radius of curvature only
depends on $Q_5$ \cite{kutasov}, \eqn{curvrad}{L^2=l_s^2 Q_5} and
hence in this case \eqn{knforf}{k_N = Q_5.} Since only the
fundamental strings end,  the radius of curvature does not jump
when crossing the AdS$_2$ brane. Instead, it is the 6d string
coupling that jumps, since it is given by
\eqn{coup}{\frac{1}{g_{s,6d}^2} = \frac{Q_5}{Q_1}.} Note that in
the D3-D5 or D1/D5 - D3 system, $q$, the number of branes ending,
was bounded by $N$, the number of branes that were there to begin
with, which in terms set the AdS curvature radius. Here $q$ is
bounded by $Q_1$, which only appears in the string coupling, not
the curvature radius. Using \coup\ the bound reads \eqn{bound}{q
\leq \frac{Q_5}{g_{s,6d}^2}} which looks like a non-perturbative
bound.

This F1/N5 -D3 system was  recently studied in  detail by Bachas
and Petropoulos \cite{bp}. This system has the advantage that the
near-horizon geometry has an exact CFT\footnote{ That is instead
of classical supergravity one can actually study tree level string
theory on this background, removing the need for small curvatures,
but still requiring small string coupling.} description as a
supersymmetric WZW model on $SL(2,{\bf R})\times SU(2)$ in which
one can find the D brane states. These turn out to be associated
with an $AdS_2\times S^2$ geometry, which is what we found as
well.

To find the branes, one first finds the branes existing in the
AdS$_3$ and $S^3$ independently, which correspond to conformal
boundary states (separately) for the $SL(2, {\bf R})$ and the
$SU(2)$. The D branes for the full geometry can be obtained by
putting these together, since the conformal boundary state for the
full theory is found by tensoring together the conformal boundary
states of each of the factors. Now the brane states of the $S^3$
are spherical $D2$-branes. They are stabilized at fixed radius by
$p$ units of magnetic flux for a system with $p$ branes, and have
radius \eqn{radiusofthesphere}{r =L \; \sin (\pi \; p \;
l_s^2/L^2).} Bachas and Petropoulos also found the D-branes of the
SL(2, {\bf R}) WZW model, which are D-strings with $AdS_2$
geometry, where the $AdS$ curvature is $l^2=L^2 (1+C^2)$ and they
find that $C$ is given by \eqn{theirc}{C=qT_F/pT_D ,} where $T_F$
and $T_D$ are the tension of the fundamental and D-strings
respectively. $q$ is the quantized $F_{tx}$ along the AdS$_2$. $p$
is the worldvolume gauge field flux through the $S^2$. Combining
the $D$-strings and $D_2$-branes together gives a single D3-brane
living in $AdS_3 \times S^3$ with associated $AdS_2\times S^2$
geometry that preserves supersymmetry and conformal invariance.
This D-brane carries the charge of a $(p,q)$. string.

Note that their result agrees with our $M=1$ result if we identify
their $q$ with our $q$ and set $p = \frac{Q_5}{2}$ with an
apropriate choice of the unknown constant of proportionality in
\c. Our $q$ is the number of strings ending on the D3. For the D1,
$q$  led to $q$ units of magnetic flux through the $S^2$. In their
case, $q$ is the number of electric flux quanta turned on along the
AdS$_2$. This may sound quite different, but it is actually once
more just counting the number of strings ending, this time
fundamental strings. But since they are turning on the S-dual
electric field, as in \cite{om}, it is the $F_{rt}$ components
that get turned on instead of the $F_{\theta \phi}$ components.

The fact that $p=\frac{Q_5}{2}$ can easily be seen by looking at the
$S^2$. Only for this particular value of $p$, the radius of the
$S^2$ \radiusofthesphere\
is indeed $L$ as we found.
$p$ counts the D1 brane charge of the wrapped D3
brane. Note that the D3 along 0123 and the NS5 along 056789 (where
6789 are the four internal directions) is dual to the NS5 012345
and D5 along 012789 of \cite{HW}, that is they are ``linked''.
What this means is that the presence of an NS5 brane induces
half a unit of D1 brane charge on the D3 brane. So in
our set-up, even though only F1 strings end on the D3 brane,
the presence of $Q_5$ NS5 branes intersecting the D3 brane still
induces
$p=\frac{Q_5}{2}$ units of D1 brane charge on the worldvolume of the
D3.
In the near horizon geometry studied by \cite{bp} any
other value of $p$ lead to a supersymmetric configuration as well.
It is unlikely that this freedom remains when including the
asymptotic flat region.

\subsection{The M2 Brane Ending on an M5 Brane}

Now let us focus on the M5 brane in the AdS$_4$ $\times$ $S^7$
geometry set up by $N$ M2 branes. For the intersecting
case, $q=0$, the story goes through as above. We end up
with a $C=0$ AdS$_3$ inside the AdS$_4$
times an equatorial $S^3$ inside the $S^7$. Once
we let $q$ of the M2 branes end, the story changes slightly.
We still turn on $q$ units of the wordvolume 3-form
flux through the $S^3$. However this
time the brane is codimension 4 and so the bending goes
like $C/r^2$, and hence
\eqn{mtwo}{ x=C/u^2}
seems to describe
the embedding inside the AdS$_4$.
It is possible that the $1/r^2$ behavior is only
adequate far away from the intersection and that in the M2 near-horizon
region the M5 brane bending actually is $1/r$ and then turns over.
This kind of behavior seems to be suggested by the analysis of
\cite{neil}, where it is found that a D-string ending on any
higher dimensional D-brane leads to a $1/r$ bending close to the brane,
while only far from the intersection one finds the naive
bending.

\subsection{Generalities}

We see that with branes ending on branes, there is in general a
spike where the branes are pulled; for example, if D3 branes end
on the 5-brane, the 5-branes grow a spike where they are pulled by
the D3 branes \cite{witten4d}. This spike could signal that the
boundary conditions actually break conformal invariance; that is
while the bulk $\beta$ function still vanishes by construction,
there might be a non-zero boundary $\beta$ function. In the case
of intersecting branes, one has the same number of  branes
 on the other brane from either side and there is no bending and
the boundary conditions clearly preserve conformal invariance. In
order to preserve conformal invariance with branes ending, the
spike has to be a $1/r$ spike.

Even though on the probe level the 4d and
2d set-ups seem to be very similar in character, it is clear that
the physics is somewhat different. While in the 4d case in the
$l_s \rightarrow 0$ decoupling limit the 6d modes decouple,
leaving us with a 4d field theory with a 3d defect, in the black
string case the decoupling limit for the 2d modes also leaves the
D3 brane gauge coupling finite. Probably in the gravity
interpretation,  this is related to the observation of \cite{bald}
that lower dimensional brane metrics never completely localize but
are always smeared.

\section{The Born-Infeld Calculation}

As a confirmation of the results derived in the last subsection
and to determine the constant $\alpha$ in \c, we study the
worldvolume action of a probe D5-brane in the AdS$_5$ $\times$ $S^5$
background set up by $N$ D3 branes.
The DBI
part of the AdS action only gets a contribution from the induced metric
and is competing with the contribution from the WZ term, which
couples the $q$ units of worldvolume flux through the sphere to the
background RR 4-form turned on along the AdS$_4$
due to the non-vanishing
RR 5-form flux turned on in the AdS$_5$.
A similar probe calculation was done for the F1/NS5 system in \cite{bp}.
There the mechanism was quite different, since it is the NSNS 2-form
that is turned on, which contributes via the DBI term of the action.
Their probe calculation confirmed the results from
the WZW calculation which we reviewed above.

So let's start with the AdS$_5$ $\times$ $S^5$ of radius $L^2 =
k_N l_s^2$. As an ansatz for the embedding we assume the D-brane
wraps an $S^2$ inside the $S^5$, but allow a representative 3rd
angle $\psi$ to be arbitrary. The induced metric on the $S^2$ is
hence \eqn{metricons2}{ ds^2 = r^2 (d \theta^2 + \sin^2(\theta) d
\phi^2) } where the radius of the $S^2$ is given by
\eqn{radiusofs2}{ r= L \sin ( \psi).} In the AdS$_5$ space ($v$,
$x$, $y$, $z$, $t$) we take the string to be static and be
embedded by \eqn{embedding}{
 v=v(x)}
so that the induced metric on this 4d part of the worldvolume
becomes
\eqn{metriconads}{ds^2 = L^2 v(x)^2 (dt^2 - dy^2 - dz^2) - L^2
\left ( v(x)^2+
\frac{v'(x)^2}{v(x)^2} \right ) dx^2.}
In addition there are $q$ units of worldvolume flux through the $S^2$
\eqn{fluxstwo}{
 \int_{S^2} F = 2 \pi q}
and since no B-field is turned on, ${\cal F} = 2 \pi l_s^2 F$.
The AdS$_5$ $\times$ $S^5$ background in addition has $N$ units
of 5-form flux turned on. We take the corresponding 4-form
vector potential to be given by
\eqn{fformvp}{
C= L^4 v^4 dx \wedge dy \wedge dz \wedge dt.}

The action of a D5 probe brane is obtained from the DBI lagrangian
\eqn{DBI}{ {\cal L} = -T_{D5} \sqrt{\det(g + 2 \pi l_s^2 F) }}
and the WZ term coupling $C$ to $F$.
First let us analyze the DBI contribution. Since $F$ is only turned
on along the sphere, $g+{\cal F}$ is block diagonal and the
determinant factorizes,
\eqn{factorize}{{\cal L} = -T_{D5} \; \;
{\cal L}_{S^2} \times {\cal L}_{AdS_4}.}
Let's first look at the $S^2$ part of the DBI
\eqn{s2part}{{\cal L}_{S^2} \; = \; \int_{S^2}  \sqrt{\det(g+{\cal F})} =
 \; 4 \pi L^2 \sqrt{\sin^4(\psi) + \frac{\pi^2 q^2}{k_N^2}}.}
Since this is the sum of two positive contributions, clearly it can be
minimized by $\psi=0$, which would correspond to a collapsed sphere,
while $\sin(\psi)=1$ is a maximum. Since we found before
that the $S^2$ actually has maximum size we proceed with
putting $\sin(\psi)=1$. Since the remaining part of spacetime
is an AdS$_4$, we don't need to be bothered by the tachyonic
mode corresponding to fluctuations in $\psi$ since they don't
violate the Breitenlohner-Freedman
bound \cite{BF}, as we will show later. For the AdS$_4$ part we just get the
contribution from the induced metric
\eqn{adspart}{ {\cal L}_{AdS_4} = \sqrt{\det(g)} = L^4
v^2 \; \sqrt{v^4  + (v')^2}.}
Last but not least we have to include the WZ part, whose
relevant piece coupling the RR 4-form to the worldvolume
field strength reads
\eqn{WZ}{{\cal L}_{WZ} =T_{D5}  \; 2 \pi l_s^2 \, F \wedge C.}
Using \fformvp\ and performing the integral over the
$S^2$ using \fluxstwo\
 this evaluates to $$T_{D5} \; 4 \pi^2 l_s^2\;  q \;  L^4 \; v^4 $$
and the effective 4d lagrangian reads \eqn{fourdlag}{{\cal L}_{4d}
= -T_{D5} \; 4 \pi \; L^4 \left ( L^2 \sqrt{1 + \frac{ \pi^2
q^2}{k_N^2} } \; v^2 \; \sqrt{v^4 + (v')^2}  - \pi\;  l_s^2\;  q
\; v^4 \right ).} Clearly, for $q=0$, we have AdS space with the
four dimensional AdS curvature 
inherited from the five dimensional one. For nonzero $q$, we still
find an AdS solution, where the two AdS curvature scales are
separated:
amazingly enough the equations of motion derived from
this lagrangian \eqn{eom}{ 2 L^2 \left ( v \sqrt{v^4 +(v')^2} +
\frac{v^5}{\sqrt{v^4+(v')^2}} \right ) - \frac{4 \pi\; l_s^2 v^3
q}{\sqrt{1 + \frac{\pi^2 q^2}{k_N^2}}} = L^2 \frac{d}{dx} \frac{v'
v^2}{\sqrt{v^4 + (v')^2}} } are indeed satisfied for
\eqn{solution}{v=\frac{C}{x}} with \eqn{finalc}{ C = \frac{\pi
q}{k_N} } in perfect agreement with \c\ for $M=1$. Using the
descent relation for D-brane tensions \eqn{descent}{T_{D(p-2)} =
T_{Dp} \; 4 \pi^2 l_s^2} this fixes $\alpha$ in \c\ in this
case to be $\frac{1}{4 \pi}$.

The same calculation goes through almost unchanged for
the D3 brane in the D1/D5 system. The effective Lagrangian for the
non-compact
2d part in AdS$_3$ similarly to above becomes
\eqn{twodlag}{{\cal L}_{2d} =
-T_{D3} \; 4 \pi \; L^2 \left ( L^2 \sqrt{1 + \frac{ \pi^2 q^2}{k_N^2} } \;
 \sqrt{v^4 + (v')^2}  - \pi\;  l_s^2\;  q \; v^2 \right )}
and one can once more verify that the equations of motion have the desired
AdS$_2$ solution.

Last but not least let's return to the issue of stability.
There seems to be a potential tachyonic instability with
the assumption that $\theta=\pi/2$, since it seems
at first glance that the brane would not be stable on the equator,
but would slip off. However, this is not the case. The scalar is
stabilized because of the surrounding AdS space. In fact,
it is easy to see the scalar satisfies the BF bound.

Let us consider a perturbation in $\psi$ so that
$\psi=\pi/2+\delta(t)$.  We evaluate det$(g +{\cal F})$, the Born-Infeld
action, where { \cal F} is the flux
\eqn{theflux}{ {\cal F} = q \pi\; l_s \; \sin\theta \;
d\theta
d \phi,}
and the metric is of the form
\eqn{linper}{
ds^2=- L^2 \sin^2 \psi \left (d \theta^2+\sin^2\theta d\Phi^2 \right)
+ L^2 {C^2 \over x^2} dt^2- \ldots }
where we only kept the time component along the AdS$_4$ direction,
which is sufficient to determine the normalization of the kinetic term.
With this we obtain
\eqn{evaldet}{\sqrt{\det (g +{\cal F})} = 4 \pi L^3 \sqrt{1 + {\pi^2q^2 \over
k_N^2} } \left ( {x \over 2 C} \dot{\delta}^2 + {C \over x}
{\delta^2 \over 1+ {\pi^2 q^2 \over k_N^2}} \right )} 
$$ \sim  {x^2 \over L^2 C^2} \dot{\delta}^2 + 2 {\delta^2 / L^2
\over 1 + {\pi^2 q^2 \over k_N^2} } $$
Recognizing ${x^2 \over L^2 C^2}$ as $g^{00}$
and
using our result that
$$l^2 = L^2 (1 + {\pi^2 q^2 \over k_N^2} )$$
we see that $\delta$
describes an excitation of
\eqn{mass}{m^2 = -{2 \over l^2} }
which is above the bound of ${- 9 \over 4 l^2}$ \cite{BF} 
independent of $q$.

\section{Beyond the Probe Approximation}
\subsection{Parameters Controlling the Backreaction}

So far we have neglected the backreaction of the D5 brane on the
background geometry of the near horizon D3 branes, or the
corresponding analogs in other dimensions. This
is important if we want to understand how
the interplay of boundary versus bulk modes in
the CFT is reflected in the spectrum of the
gravity fluctuations.  Since at the moment the
full localized supergravity solution is not known, we can't
present an analytic formula for the exact warpfactor. Given our
knowledge about the symmetries and asymptotic behavior it
shouldn't be too difficult to construct it along the lines of
\cite{smith,arvind}. At least as interesting would be to construct
the 5d supergravity solution that would lift to our solution. A
future goal it to  construct the solution as an AdS domain wall in
${\cal N}=8$ gauged supergravity in 5d, since effectively 5
dimensions got compactified in the D3 brane throat. The internal
geometry of the $S^5$ will be reflected in a flow of the 5 $SO(3)
\times SO(3)$ invariant scalars in the theory.

Let us briefly consider some general scaling arguments. In the
5d string frame the Einstein-Hilbert term is given by
\eqn{einsteinhilbert}{ \frac{L^5}{g_s^2} R } while the tension of
$M$ D5 branes wrapping the $S^2$ of radius $L^2$ with $q$ units of
flux is given by \eqn{effectivetension}{4 \pi \; M \;  T_{D5} (L^2 \;
\sqrt{1+ \frac{\pi^2 q^2} {k_N^2}} )} where we need to recall that
$T_{D5} \sim \frac{1}{g_s}$. So the jump in extrinsic curvature
is proportional to \eqn{source}{ M g_s \sqrt{1+ \frac{\pi^2
q^2} {k_N^2}}. } First note that for $g_s \rightarrow 0$, $g_s N$
fixed, but $M g_s q$, $g_s M \rightarrow 0$, the backreaction can
be neglected. In this regime the probe calculation is accurate; we
get a very flat AdS$_4$ (note that it is $q$ and not $g_s q$ that
controls the curvature of the AdS$_4$), but of course we won't get
localized modes. For this we do need to consider a parameter
regime with a significant  backreaction.

There are two interesting cases, both of which have a free
parameter that permits us to separate the four and five
dimensional curvature scales. In this paper, we consider large
$g_s q$ with finite $M$ and vanishing $g_s M$. As the probe
calculation of the previous section showed, by adjusting $q$ at
$g_s q=0$ we can start with an arbitrarily flat AdS$_4$ and then
continously turn on the backreaction. Since $g_s M \rightarrow 0$
we won't source the dilaton or the 2-form. The 5-brane carrying
3-brane charge gets replaced with $q$ smeared D3 branes. The
second case is  considered in Ref. \cite{short} and is $q=0$ and
finite $g_s M$, and we expect can localize gravity.

\subsection{Possible Behavior of the Warp Factor}

Before we discuss the gravity solutions with the backreaction
included,  let us pause and discuss the two possible behaviors of
the warpfactor and what this would mean in
terms of the dual CFT. For this we study as a toy model a thin,
charged 3-brane in AdS$_5$. That is we allow the AdS$_5$ radius to
jump from $k_N^2 l_s^2$ on the left to $k_{N-q}^2 l_s^2$ on the
right.

\begin{figure}
   \centerline{\psfig{figure=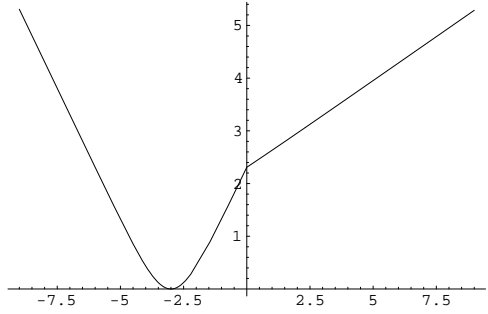,width=2.5in}}
    \caption{ ``Up-slower up'' AdS warpfactor for $L_L=1$ and $L_R=3$
.}
\label{aupupads}
 \end{figure}

\begin{figure}
   \centerline{\psfig{figure=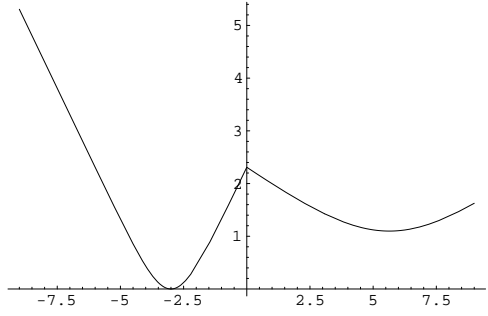,width=2.5in}}
    \caption{ ``Up-down'' AdS warpfactor for $L_L=1$ and $L_R=3$
.}
\label{aupdoads}
 \end{figure}

Turning on the backreaction, the warpfactor has to jump at the
brane. In a $Z_2$ symmetric set-up as in \cite{RS,kr}, this has to
be ``up-down''. That is, if the warp factor initially has positive
slope, on the other side of the positive tension brane it has
negative slope.  For the
asymmetric case with the additional jump in background
cosmological constant, there are in principle two possibilities,
which we refer to as ``up - slower up'', Fig.\ref{aupupads}, and
``up-down'', Fig.\ref{aupdoads}. For a thin brane of tension
$\lambda$ the warp factors read \eqn{warpfactors}{e^A = \left \{
\begin{array}{ll} \sqrt{|\Lambda|} L_L \cosh (b_L + k_L r) \; \mbox{
   for  } \; r<0 \\
 \sqrt{|\Lambda|} L_R \cosh (b_R + k_R r) \; \mbox{
   for  } \; r>0
\end{array} \right . }
where $|k_{L,R}| =  \frac{1}{L_{L,R}}$, $k_L>0$ and the sign of
$k_R$ distinguishes the two possibilities. The parameters
$b_{L,R}$ are determined from the jump conditions \eqn{jump}{L_L
\cosh(b_L) = L_R \cosh(b_R) , \; \; \; \; \lambda=\frac{3}{2} (k_L
\tanh(b_L) - k_R \tanh(b_R) ) } and the 4d cosmological constant
of the model is \eqn{fourdcc}{\Lambda=\frac{1}{9 \lambda^2} \left
[ (\frac{9}{4} (k_L - k_R)^2 - \lambda^2 ) \; (\frac{9}{4} (k_L +
k_R)^2 - \lambda^2 ) \right ] .}

\begin{figure}
   \centerline{\psfig{figure=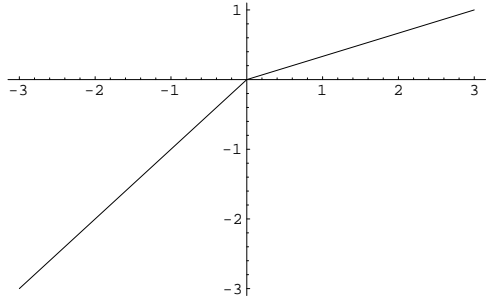,width=2.5in}}
    \caption{ ``Up-slower up'' Minkowski warpfactor for $L_L=1$ and $L_R=3$
.}
\label{aupup}
 \end{figure}

The difference between the two in terms of the gravity fluctuations
is that the ``up-down'' warpfactor has a mode that
approximately gives rise to a 4d Newton's law, while the
``up-slower up'' warpfactor always looks five dimensional.
In order to see this, consider for simplicity the ``up - slower up''
Minkowski set-up, as it would for example arise on the Higgs branch
of the ${\cal N}=4$ SYM, where $SU(N+1)$ is broken to $SU(N)$.
The additional D3 brane that is separated from the stack of $N$
D3 branes in the AdS$_5$ geometry appears as a domain wall, across
which the background curvature jumps.
\begin{figure}
   \centerline{\psfig{figure=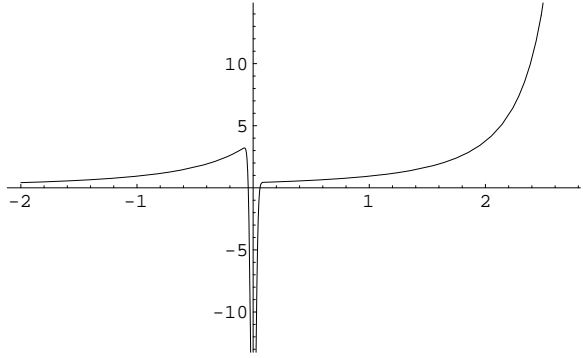,width=3.0in}}
    \caption{Volcano potential associated to the ``up-slower up''
Minkowski warpfactor for $L_L=1$ and $L_R=3$
.}
\label{vupup}
 \end{figure}
Fig.\ref{vupup} displays the volcano potential in this case.
Obviously this does not lead to localized gravity. The zero mode
is not normalizable, since the warp factor diverges on the right.
However, as in the AdS case, this would lead to 4d physics if there
were a bound state (or in the Minkowski case a resonance)
that dominated over the KK modes.  This clearly
does not happen in this case, where on the right
hand side, the bound state
amplitude would be small, while the KK modes are unsuppressed,
though coming in  from the left, they  are  surpressed by the
usual barrier.  The barrier on the right is set by $1/L_R^2$ (the
onset of the potential on the right of the delta), which is also
the natural mass scale for the resonances, so there is no regime
in which the 5d modes can be separated from a ``4d'' resonance. In
the AdS$_4$ warping all modes are normalizable, but again to have
the light mode separated from the 5d KK modes, we need an
``up-down'' warpfactor.

It is interesting to understand the distinction between the
localizing warp factors in terms of the dual CFT. After all, both
situations correspond to an AdS brane (in the geometries we study)
with the dual CFT living on half the boundary, as previously
described.  We have argued that in the ``up-down'' situation, we can
get a 4d Newton law, while in ``up-slower up'' Newton's law
essentially stays 5d. In the dual CFT this has to be reflected in
a property of the 2-point function of the stress-energy tensor. The
bulk 2-point function goes like $\frac{1}{|x|^8}$, consistent with
coupling to the 5d graviton. Since the presence of the boundary
broke the symmetry to $SO(3,2)$ already, we can have a 3d CFT
living on the codimension one defect coupled to the 4d CFT. Its
stress energy tensor to leading order
has $\frac{1}{|x|^6}$ correlations and hence
naturally couples to a 4d graviton. So the difference between
``up-down'' and ``up-slower up'' can be interpreted as  whether we
get $\frac{1}{|x|^8}$ or at least approximately $\frac{1}{|x|^6}$
correlators.
For latter it is necessary but not sufficient to
have a dynamical 3d CFT on the defect.

\subsection{Turning on the Backreaction}

\begin{figure}
   \centerline{\psfig{figure=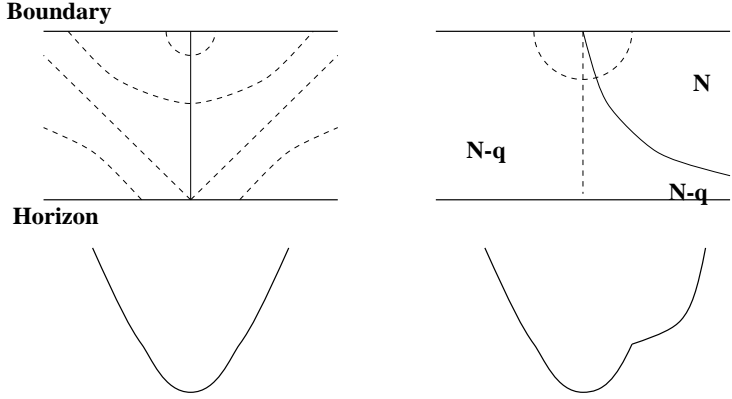,width=3.8in}}
    \caption{The $q=0$ and finite $q$ D5 branes in the
D3 brane near horizon geometry. The dashed lines show the
lines parametrized by the $r$ coordinate of the AdS$_4$ slicing.
The corresponding warp factors as a function of $r$ are also displayed.}
\label{bendingfig}
 \end{figure}

Let us first discuss the case with finite $M$, so that $g_s M \rightarrow 0$
and $g_s q$ finite. In this case the AdS$_4$ can be very flat
already in the probe limit.
In addition only 3-brane fields are turned on, the source terms
for dilaton and 2-form vanish, so that the analysis
of the supergravity solution should simplify.
However we will argue that there are several reasons why
this warp factor has to be ``up-slower up''.

First consider small $g_s q$. Up-down is the only choice that
continously reproduces the probe result as $g_s q$ goes to zero.
In the probe limit the warpfactor is just the pure AdS$_5$ warpfactor
\eqn{adswarp}{e^A(r) = \sqrt{|\Lambda}| L \; \cosh (r/L-b) }
with the brane at $r=0$ sitting at a warpfactor
$\sqrt{|\Lambda}| L \; \cosh(b)$, which is large for large $q$.
Now we want to turn on the backreaction. The warpfactor
becomes different on both sides and is characterized by $b_L$ and $b_R$
respectively. $b$ gives the distance between brane and turnaround point.
The sign of $b$ tells us if the turnaround is to the left or to
the right of the brane.
Writing the pure AdS warpfactor (\ref{adswarp}) from the
probe limit as a left and a right warpfactor, we have $b_L=b_R=b$.
There is only one turnaround point and
it is to the left of the brane.
Now as we smoothly turn on $g_s q$, $b_L$ and $b_R$ will
separate proportional to $g_s q$. For an ``up down''
warpfactor we need a second turnaround point to
the right of the brane, that is we need $b_R$ to change sign.
Since we start out with large $b$, changing the sign is not a small
peturbation. This argument rules out ``up-down''
for small $g_s q$, we will now present two more
arguments that suggest that this is still the case once
we turn on a significant backreaction.

A picture of this situation as seen in the AdS$_5$ Poincare
patch (as it arises in the near horizon limit of the D3 branes) is
displayed in Fig. \ref{bendingfig}.
Far to the right (for large $x_6$) the
5 brane is almost horizontal, so that locally we just get the ${\cal N}=4$
SYM on the Higgs branch, breaking $SU(N)$ to $SU(N) \times SU(N-q)$.
\footnote{This picture actually seems to capture quite acurately what
is going on in the dual SYM. In order to get the situation with $SU(N)$
on one side of the defect and $SU(N-q)$ on the other side of the defect,
one can start with $SU(N)$ everywhere and turn on a VEV
for one of the impurity hypermultiplets from the 3-5 strings,
corresponding to moving off $q$ D3 branes on the left \cite{sethi}.
The vacuum solution in the presence of this VEV seems to demand that
on the right now the bulk Higgs fields have to develop a
position dependend $\frac{1}{x_6}$ VEV, so that the $SU(N)$
on the right gets broken to $SU(N-q)$ as well and is only restored
asymptotically. This is how string
theory realizes the interaction of two conformal
field theories with different central charge across a boundary.
We like to thank E. Martinec for a useful discussion
on this point.} For this case the full 10d metric is known, since it
is just a multi-centered D3 brane metric. By inspection one finds
that the warp factor in this case is ``up-slower up'',
e.g. in 5d language, an $SO(6)$
invariant configuration of D3 branes gives rise to the
warpfactor of Fig. \ref{aupup}.
By continuity we expect this then to be the case everywhere
along the brane, not just at large $x_6$.

Last but not least,
the dual CFT also supports the point of view that the $g_s M=0$,
$g_s q$ finite configuration does not localize gravity. Above we
argued that 3d CFT living
on the boundary defect is required. For a single D5 brane intersecting
the $N$ D3 branes, the only 3d degrees of freedom are the hypers
from 3-5 strings, which should flow in the IR to a theory containing
no dynamical degrees of freedom.

In another paper, we  consider the second possibility, that is
turning
on $g_s M$ while leaving $q=0$.
In this case we have an additional $Z_2$ symmetry,
so that ``up-slower up'' is not a possibility. The brane
starts out sitting at the turnaround point of the warpfactor
in the probe limit. Including its
backreaction by turning on $g_s M$ the warp factor
will this time grow a little ``up-down'' spike in the center, as we
discuss in more detail in \cite{short}.
In this case, all three of the above arguments fail and
we expect the theory does in fact have a parameter regime
in which gravity is localized.



\section{Conclusions}

In this paper we have given a stringy realization of the
duality conjecture in \cite{kr} between conformal
field theories on a manifold with boundary and an AdS brane
inside AdS. We showed that in brane configurations
which realize such supersymmetric CFTs with codimension
one conformal defects, the higher dimensional
brane in the near horizon geometry of the lower dimensional
brane spans a worldvolume of the form AdS times sphere.
We gave some general arguments of how the 3 numbers $M$, $N$ and
$q$ for $q$ out of $N$ lower dimensional branes ending on
$M$ higher dimensional branes, control the backreaction
and hence the ratio of 4d and 5d curvature scales (refering
to the D3 D5 system to which we devoted most of our attention)
and how
the interplay of 4d and 5d modes in the gravity reflects
the interaction between bulk and boundary modes in the CFT.

\section*{Acknowledgements}
We are indebted to Amihay Hanany for suggesting that the 3-brane
5-brane system is a natural place to look for a stringy
realization of a CFT with boundary. We would like to thank Andy
Strominger for many useful suggestions and observations. We would
also like to thank Emanuel Katz, Emil Martinec, Rob Myers,
Amanda Peet and Sav
Sethi.  Further we'd like to thank Oliver DeWolfe, Dan Freedman
and Hiroshi Ooguri for pointing out a missing factor of 2 in eq. (42).

\bibliography{3dcft}
\bibliographystyle{ssg}
\end{document}

%% file: lmacros.tex
\input umacros
\def\lbldef#1#2{\expandafter\gdef\csname #1\endcsname {#2}}
\def\eqn#1#2{\lbldef{#1}{(\ref{#1})}%
\begin{equation} #2 \label{#1} \end{equation}}


%% file: umacros.tex
\def\TL{\hfil$\displaystyle{##}$}

\def\TT{\hbox{##}}
\def\seqalign#1#2{\vcenter{\openup1\jot
  \halign{\strut #1\cr #2 \cr}}}

\def\comment#1{}
\def\fixit#1{}

\def\overleftrightarrow#1{\vbox{\ialign{##\crcr
     $\leftrightarrow$\crcr\noalign{\kern-0pt\nointerlineskip}
     $\hfil\displaystyle{#1}\hfil$\crcr}}}

\def\lsim{\mathrel{\mathstrut\smash{\ooalign{\raise2.5pt\hbox{$<$}\cr\lower2.5pt\hbox{$\sim$}}}}}
\def\gsim{\mathrel{\mathstrut\smash{\ooalign{\raise2.5pt\hbox{$>$}\cr\lower2.5pt\hbox{$\sim$}}}}}

\def\sqr#1#2{{\vcenter{\vbox{\hrule height.#2pt
         \hbox{\vrule width.#2pt height#1pt \kern#1pt
            \vrule width.#2pt}
         \hrule height.#2pt}}}}

\def\href#1#2{#2}  
